\documentclass[useAMS,usenatbib]{mn2e}
\usepackage{times}
\usepackage{amssymb}

\newcommand{\kms}{km~s$^{-1}$}

\usepackage{graphics,epsfig}
\title[NGC~3741]{Life in the last lane: Star formation and chemical evolution in an extremely gas-rich dwarf}
\author[]
{
Ayesha Begum$^{1}$\thanks{E-mail:ayesha@ast.cam.ac.uk},
Jayaram. N. Chengalur$^{2}$,
Robert C. Kennicutt$^{1}$,
Igor D. Karachentsev$^{3}$ \&
\newauthor Janice C. Lee$^{4,5}$ 
\\
\\
$^{1}$Institute of Astronomy, University of Cambridge, Madingley Road, Cambridge, CB3 0HA, UK\\
$^{2}$National Centre for Radio Astrophysics, Post Bag 3, Ganeshkhind, Pune 411
007, India\\ 
$^{3}$Special Astrophysical Observatory, Nizhnii Arkhys 369167, Russia\\
$^{4}$Observatories of the Carnegie Institution of Washington, 813 Santa Barbara Street, Pasadena, CA 91101\\
$^{5}$Hubble Fellow\\
}
\begin{document}
\date{}
\maketitle
\begin{abstract}
We present an analysis of HI, H$\alpha$,  and oxygen abundance data for NGC~3741. 
This galaxy has a very extended  gas disk ($\sim$ 8.8 times the Holmberg radius), 
and a dark to luminous (i.e. stellar) mass ratio of $\sim 149$, which
makes it one of the ``darkest'' dwarf irregular galaxies known. 
However its ratio of baryons (i.e. gas + stellar) mass to dark
mass is typical of that in galaxies. Our new high-resolution 
HI images of the galaxy show
evidence for a large scale (purely gaseous) spiral arm and central bar.
From our HI data, a rotation curve can be derived out to 
$\sim$ 37$-$ 44 disk scale lengths in the J and B band respectively. This is
just slightly short of the radius at which one would expect
an NFW type rotation curve to start falling. 
The galaxy has an integrated star formation
rate of $\sim$ 0.0034 M$_\odot$~yr$^{-1}$, while the
average star formation rate within the optical disk is 
$\sim$ 0.0049 M$_\odot$~yr$^{-1}$~kpc$^{-2}$. 
Despite the gaseous spiral feature and the 
on-going star formation, we find that the global gas density in NGC~3741 is
significantly lower than the Toomre instability criterion.
This is consistent with the behaviour seen in other dwarf galaxies.
We also find that the star formation rate 
is consistent with that expected from the observed
correlations between HI mass and SFR and the global Kennicutt-Schmidt law
respectively. We measure the oxygen abundance to be 12+ log(O/H)=7.66$\pm$0.10, 
which is consistent with that expected from the
metallicity-luminosity relation, despite its extreme gas mass ratio.
We also examine the issue of chemical evolution of
NGC~3741 in the context of closed-box model of chemical evolution. 
The effective oxygen yield of NGC~3741 is consistent with
recent model estimates of closed-box yields, provided one
assumes that the gas has been efficiently mixed all the way
to edge of the HI disk (i.e. $> 8$ times the optical radius).
This seems a priori unlikely. On the other hand, 
using a sample of galaxies with both interferometric HI maps and chemical
abundance measurements, we find that the effective yield is anti-correlated with the total
dynamical mass, as expected in leaky box models.

\end{abstract}
\section{Introduction}
\label{sec:intro}

The processes that govern the conversion of gas to stars in galaxies are complicated and poorly understood. Nonetheless, there do exist 
empirical star formation recipes which are widely applied to models of galaxy evolution. Typically, these recipes allow one to compute
a global star formation rate given the gas column density or, alternatively, some estimate of the dynamical timescale (e.g. the angular rotation frequency $\omega$). While the existence of such empirical relationships is fairly well established in large galaxies (e.g. Kennicutt 1998), it is unclear whether these are applicable to the smallest dwarf galaxies (e.g. \citealp{begum06}). However, star formation in such small galaxies is of particular interest in studies of galaxy formation, because, in hierarchical models of galaxy formation, small galaxies form first and then merge to form larger galaxies. Nearby, metal poor dwarf galaxies are, in this sense, 
analogs of first units of star formation in the universe.  Detailed studies (at resolutions that are not achievable for high
redshift objects) of the star formation  in nearby dwarf galaxies could hence give some insight into the star formation processes occurring in primordial galaxies at high redshifts.

\begin{figure*}
\psfig{file=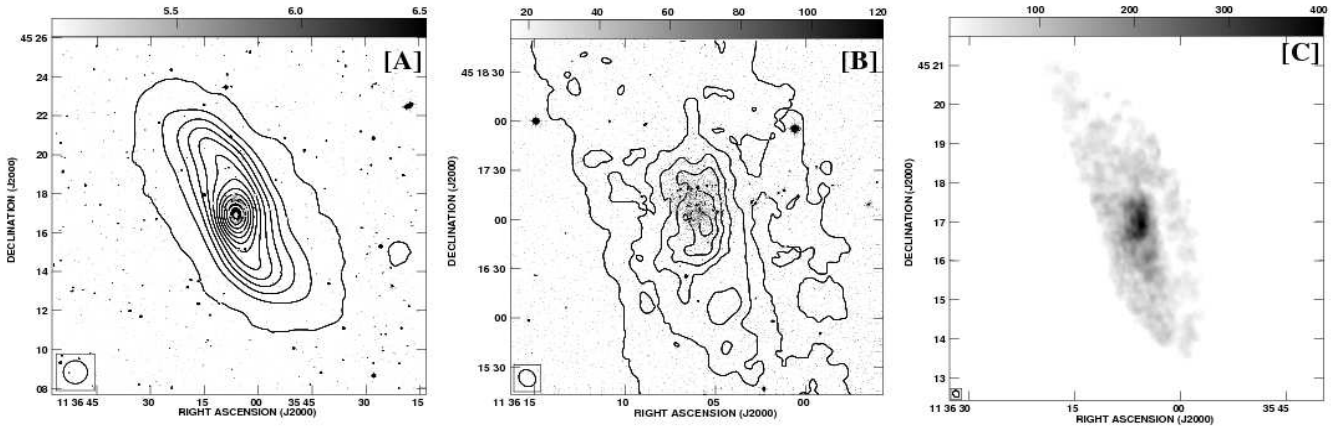,width=7.0truein}
\caption{{\bf{[A]}}The optical DSS image of NGC~3741 (greyscale) with 71$^{''}\times71^{''}$ resolution integrated
          HI column density map (contours) overlayed. The contour levels are 0.1, 1.34, 2.59, 3.83, 5.08, 6.32, 
7.57, 8.82, 10.05, 11.31, 13.79 and 15.04 $\times 10^{20}$ cm$^{-2}$.
{\bf{[B]}} J band image of NGC~3741 (grey scale) overlayed on  HI column density map (contours) at 
11$^{''}\times9^{''}$ resolution. The contour levels are 2.8,10.73, 18.66, 16.57, 34.5, 42.4  
$\times 10^{20}$ cm$^{-2}$.
{\bf{[C]}}Greyscale HI image of NGC~3741 at 11$^{''}\times9^{''}$ resolution
. The range of grey scale  is from 2.5$\times~10^{19}$ cm$^{-2}$ to 4.2 $\times~10^{21}$ cm$^{-2}$.
The image shows one sided spiral arm emanating from a central bar like region on the western side of the galaxy.
}
\label{fig:overlay}
\end{figure*}

We have been conducting a systematic multi-wavelength survey
(the Faint Irregular Galaxy GMRT Survey -- FIGGS (\citealp{begum06})
of nearby dwarf galaxies. This survey revealed
the truly unusual properties of NGC~3741, with an HI disk
that extends to more than $\sim$8 times its Holmberg radius
(\citealp{begum05}, henceforth Paper~I).  NGC~3741 has the 
most extended gas disk known. Further, with the observed M${\rm{_{HI}/L_B=5.8\pm1.4}}$, 
NGC~3741 is also 
one of the most gas-rich dwarf galaxies known. In Paper~I we presented a rotation curve
for NGC~3741 that extended out to $\sim 38$ optical B band scale lengths (Bremnes et al. 2000) and showed that 
within the last measured point of the rotation curve, the dark matter in NGC~3741 
contributes to 92 \% of its total dynamical mass, whereas stars contribute only 1\% of the total mass. On the
other hand, we also found that its total baryonic (i.e. HI + stellar mass) to dark matter ratio of
${\rm{M_{bar}/M_T}}\sim0.08$ falls within the range that is typical
of galaxies ($\sim0.07-1.0$; Paper~I). 
NGC~3741 thus seems to be a case of inefficient star formation, that is worthy of further study.

    In this paper we present new HI and H$\alpha$ data for NGC~3741. We also estimate the gas-phase oxygen
abundance of this galaxy using the spectrophotometric  data from \cite{mou06}.
We use these data to study the the relationship between gas and star formation  and  abundance properties of 
this very gas-rich galaxy.
Throughout the paper, we adopt a distance of 3.03 Mpc, as derived from the tip of the red giant
branch (Karachentsev et al. 2003). At this distance 1$^{\prime \prime}$ corresponds to 14.7 pc.

\section{Observations and data reduction}
\label{sec:obs}

\subsection{HI Observations}
\label{ssec:hi}

\begin{figure*}
\psfig{file=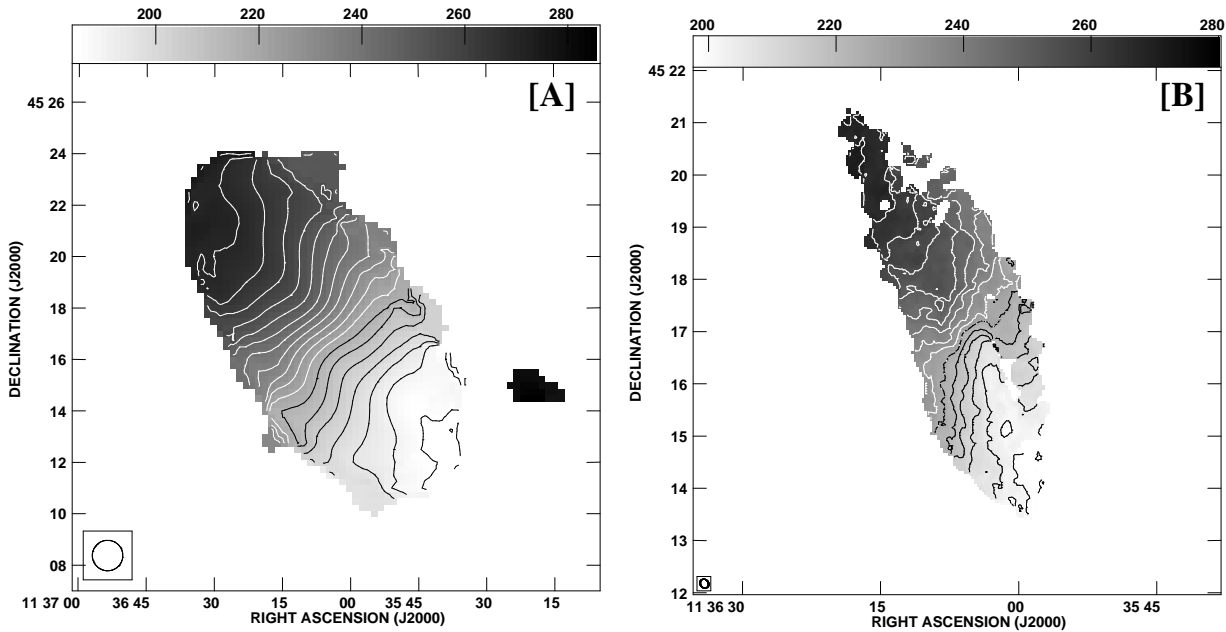,width=6.5truein}
\caption{{\bf{[A]}}The HI velocity field of NGC~3741 at 71$^{''}\times71^{''}$resolution. 
{\bf{[B]}} The velocity field at 11$^{''}\times 9^{''}$ resolution obtained from the combined data.
	The contours in both images are in steps of 5 \kms and range from 180 \kms to 275 \kms.
}
\label{fig:mom1}
\end{figure*}

     HI observations of NGC~3741 were conducted at the GMRT (Giant Metrewave Radio Telescope), DRAO (Dominion 
Radio Astrophysical Observatory) Synthesis Telescope 
and WSRT (Westerbork Synthesis Radio Telescope). A combination of the data from these arrays provides sensitivity to the
large-scale, low surface brightness component which could have been missed 
with the GMRT observations alone, because of missing short spacings at the GMRT. 
The GMRT observations have been described in detail in Paper~I. 
Here we only note that the final visibility set had 128 spectral channels spread across a 1 MHz bandwidth, i.e. a  channel spacing of $\sim 1.6$~kms$^{-1}$. The DRAO observations were made in 
a "complete survey" mode (Landecker (2000)). Twelve sets of spacings were used, giving baselines from 12.9 to 604.3 m at 4.3 m intervals. The DRAO data contained 256 spectral  channels across a 1 MHz bandwidth, 
or a channel separation  of $\sim 0.8$~\kms. The WSRT observations consisted of 
a single 12hr observing run in the ``maxi-short'' configuration taken on 09$^{th}$ Oct, 2005. There 
were a total of 2048 spectral channels across a 10 MHz bandwidth, for a channel 
separation of $\sim 1.0$~\kms. The flux calibration was performed 
using an observations of 3C147, while phase calibration was done using self-calibration on the 
continuum sources in the field.

    A low resolution $\sim 70^{''}$ image was made from combining the GMRT, WSRT  and DRAO data. 
This image was made as follows. After continuum subtraction (using AIPS (Astronomical Image
Processing Systems) task UVSUB ) the GMRT, DRAO and WSRT visibility data were exported to MIRIAD 
where they were regridded onto a common velocity scale using the task ``uvaver''. 
The channel width after regridding was 2.0 \kms. The regridded visibility data
 were read back into AIPS and CLEAN cubes were made from each data set using the task IMAGR. 
The uv range and taper were adjusted for each data set, so as to give a circular 
beam of $\sim 71''$ resolution. 
The CLEAN cubes were then corrected for  the primary beam attenuation appropriate 
for each telescope using the task PBCOR in AIPS. These primary beam corrected cubes were 
then combined (using the weights determined by the rms noise in each cube)  to 
a final combined data cube. Since the rms noises in the DRAO, GMRT and WSRT cubes was 13.6 mJy, 5.5 mJy and 
0.96 mJy respectively, the weighted sum is dominated by the WSRT data.  Emission in the combined cube is seen  over the velocity range 166 $-$ 293.7 \kms. In contrast, in the GMRT data alone, emission is detectable only over the range 180.0 $-$ 283.7 \kms.  In Paper~I we only presented a $\sim 40^{''}$ resolution image  made from the GMRT data. However since the GMRT  has a hybrid configuration, higher resolution images can also be made from the same data. Here we present both  $\sim 10^{''}$ and $\sim 4^{''}$ resolution images made from the same data.  The rms noise in these data cubes is 1.6 mJy and 1.1 mJy respectively. 

\subsection{H$\alpha$ Observations}
\label{ssec:halpha}
Narrow-band H$\alpha$ imaging of NGC~3741 was obtained as part
of the 11 Mpc H$\alpha$ and Ultraviolet Galaxy Survey (11HUGS; Kennicutt et al. 2005, Lee 2006).
NGC~3741 was observed on 2001 March 31, using the
2K CCD imager on the Steward Observatory Bok 2.3~m telescope.
The onband image was taken through a 3-cavity interference
filter with central wavelength 658.5~nm and FWHM 6.6~nm,
with an integration time of 1000s.  For continuum subtraction
the galaxy was observed with a Kron-Cousins $R$ filter, with
an integration time of 200~s.  Details of the observation
and data reduction procedures can be found in Lee (2006) and
Kennicutt et al. (2007, in preparation).

Raw images were reduced following standard procedures using
IRAF\footnote{The Image Reduction and Analysis Facility (IRAF) is
distributed by the National Optical Astronomy Observatories, which are
operated by AURA, Inc.\ under cooperative agreement with the National
Science Foundation.}.  Astrometric alignment and calibration was
performed using reference stars from the USNO-A2 catalog, and
flux calibration was performed using observations of
spectrophotometric standard stars from Massey et al. (1988).

The processed narrow-band images contain contributions both from
H$\alpha$ and [NII] line emission as well as underlying stellar
continuum (including H$\alpha$ absorption).  Net emission-line images
were obtained by subtracting a scaled $R$ image from the narrow-band
image,
aligning the respective images using foreground stars.  Additional
small corrections for line emission in the $R$-band image and the
wavelength shift between filters were applied, as described in
Lee (2006) and Kennicutt et al. (2007).  Finally a small correction
for [NII] contamination in the image was applied using an integrated
spectrum of the galaxy obtained by Moustakas \& Kennicutt (2006).
The derived total H$\alpha$ luminosity for NGC~3741 is 
$\sim 4.3 \times 10^{38}$ erg s$^{-1}$, which corresponds to 
a star formation rate (SFR) of 3.4$\times 10^{-3}$ M$_\odot$yr$^{-1}$ 
(using the conversion factor from \citealp{kennicutt98}).

\section{Results}
\label{sec:result}

\subsection{HI distribution and kinematics}
\label{ssec:HIdistribution}


The integrated flux measured from the $71^{''}$ resolution data cube (see Fig.~\ref{fig:overlay}[A]) is $59.6$~Jy~\kms. The HI mass
obtained from the integrated profile  is ${\rm{1.30 \times{10}^{8}M_\odot}}$, and the ${\rm{M_{HI}/L_{B}}}$  ratio is found to be $\sim 4.7$ in solar units. This estimate of the  HI mass (which is dominated by the WSRT data) is slightly ($\sim 1.5\sigma$) lower than the value of ${\rm{1.6\pm0.2 \times{10}^{8}M_\odot}}$ that was obtained in Paper~I.   
 On comparing Fig.~\ref{fig:overlay} with the  $\sim 40^{''}$  map made using only the GMRT data (in Figure 1[A] of Paper~I), we find that  using the combined data set from WSRT, GMRT and DRAO, we could recover the faint, more extended   outer envelope of HI distribution. The HI envelope extends to $\sim14.6^{'}$, at a level of $\sim 1 \times 10^{19}$ cm$^{-2}$ i.e. $\sim$~8.8 times the Holmberg
diameter, compared to $\sim8.3$ times the Holmberg diameter measured with the GMRT data alone. 

  In Paper~I we had noted that NGC~3741 seemed to have a central bar, but this is 
not apparent in the lower resolution image in Fig.~\ref{fig:overlay}. On the other 
hand, the 10$^{''}$ resolution image (Fig.~\ref{fig:overlay}[C]) 
shows the central bar clearly, and also shows a large one sided spiral arm of $\sim$9$^\prime$ 
(with no stellar counter part) on the western side of the galaxy, emanating 
from a central bar like region. A presence  of an HI bar is also evident in 
the high resolution velocity field (Figure~\ref{fig:mom1}[B] here  and Figure.1[B] in Paper~I) 
from the oval distortion of the isovelocity contours in the center of the galaxy.
Further, Gentile et al.(2007) also found evidence for a bar at the centre of this galaxy based 
on a harmonic decomposition of the velocity field. On the other hand,  near infra-red images of this 
galaxy shows no evidence for a stellar bar. Figure.~\ref{fig:overlay}[B] shows the 
10$^{''}$ resolution HI distribution (contours) overlayed on the J band emission (greyscale) from
Vaduvescu et al.(2005). Vaduvescu et al.(2005) estimated an ellipticity and PA of the near infra-red
disk emission to be 0.26 and 23 deg respectively. On the other hand, PA and ellipticity of the HI bar
derived from 10$^{''}$ resolution HI map is $-$10$\pm$5 deg and 0.65$\pm0.10$.
Thus, while the optical morphology of NGC~3741 implies a disk structure (Vaduvescu et al.(2005)), 
our HI data suggest a bar configuration in the center of the galaxy. Such difference in the optical
and HI morphology has been noted previously in a blue compact dwarf galaxy NGC 2915 (Meurer et al. 1996).

 The velocity field of NGC~3741, as  derived from a moment analysis
of 71$''\times 71''$ resolution data cube, is shown in Fig.~\ref{fig:mom1}[A]. 
A warp is seen 
in the outer regions of the galaxy, as apparent from the bending of the iso-velocity contours
(Figure.~\ref{fig:mom1}).

\subsection{HI rotation curve}
\label{ssec:rotcur}

\begin{figure}
\psfig{file=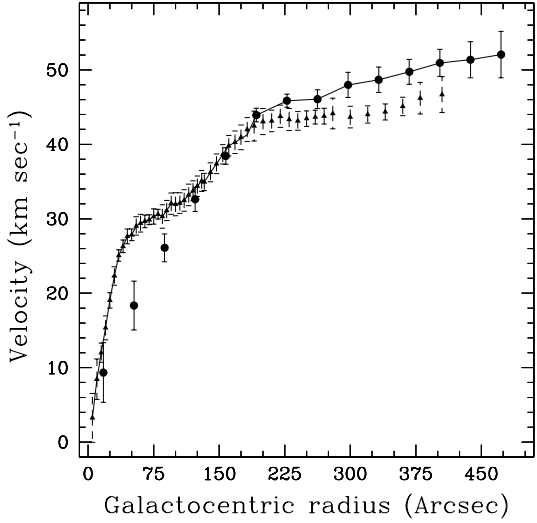,width=3.0truein,angle=0}
\caption{The rotation curve for NGC~3741 derived from the velocity field  at 71$^{''}\times71^{''}$ resolution 
	(solid points) and from the GMRT velocity fields at various resolutions (triangles). The adopted `hybrid' rotation 
	curve is shown as a solid line.
}
\label{fig:vrot}
\end{figure}

The rotation curve derived from the GMRT data was presented in Paper~I. However,
 as our combined data reveals a more extended HI disk, we have
re-derived the rotation curve for NGC~3741 from 71$''\times 71''$ resolution 
velocity field, using the tilted ring model.
The derived rotation 
curve is shown in Fig.~\ref{fig:vrot} as solid circles. Also shown in the 
figure is the rotation curve derived in Paper~I (i.e. from the high resolution 
GMRT data alone). As can be seen, in the inner region of the galaxy, the 
rotation curve derived from the high resolution data is significantly steeper, 
most likely because of beam smearing in the low resolution image.  
The `hybrid' rotation curve shown as a solid line in Fig.~\ref{fig:vrot} 
consists of the rotation velocities derived from the high resolution 
GMRT data in the inner regions,  and  from the low resolution velocity 
field from the combined data in the outer regions.  The  ``asymmetric drift correction" 
was found to be small compared to the errorbars 
at all radii and hence was ignored.  The final adopted rotation curve is measured 
up to a record $\sim$ 37 $-$ 44 disk scale lengths in 
J band  and B band, respectively (Vaduvescu et al. 2005, Bremnes et al. 2000).

As noted in Paper~I and in Section~\ref{ssec:HIdistribution}, the velocity field
of NGC~3741 shows clear signs of non-circular motions. These have
not been accounted for in our derivation of the rotation curve.
Gentile et al. (2007) used the WSRT data to develop a kinematical model for NGC~3741
to account for the non-circular motions. We find that 
our rotation curve is systematically higher that Gentile et al. (2007) 
in the inner regions of the galaxy, but the two curves match within one sigma 
errorbars. For the remainder of our analysis we have used the `hybrid'  rotation 
curve in Fig.~\ref{fig:vrot}, though using the rotation curve by Gentile et al. (2007)
does not make any significant difference to our results. 
The parameters for NGC~3741 derived from the last measured point of the rotation curve,
along with other parameters are listed in Table~\ref{tab:par}. 
The stellar mass to light ratio in the B band, $\Gamma_B$ was 
derived from the observed colour (B-V)$\sim$0.36 (Taylor et al. 2005), using from the
low metallicity Bruzual $\&$ Charlot SPS model from Bell $\&$ de Jong (2001).
As discussed in Paper~I, there is considerable uncertainty in 
the mass modelling of NGC~3741 due to the presence of an HI bar 
in the galaxy, hence $\Gamma_B$ obtained from the mass modelling 
was not considered for obtaining the stellar mass. 
With 
M$_{\rm T}$/L$_{\rm B}$ $\sim 149$, NGC~3741
is  one of the ``darkest'' known gas-rich galaxies.It is interesting to note that NGC 3741 belongs to a dwarf
galaxy association which is also very dark matter dominant
(assuming that the association is bound) (Tully et al. 2006). 

\begin{table}
\caption{Derived parameters for NGC~3741}
\label{tab:par}
\vskip 0.1in
\begin{tabular}{ll}
\hline
Parameters& Value \\
\hline
\hline
${\rm{M_{gas}}~(1.3M_{HI})}$& 1.69 $\times~10^8$~M$_\odot$\\
${\rm{L_B}}$& 2.7 $\times~10^7$~L$_\odot$\\
${\rm{M_{gas}/L_B}}$& 6.26\\
Mass-to-light ratio, $\Gamma_B$&0.51 \\
Stellar mass ${\rm{M_*}}$& 1.38 $\times~10^7$~M$_\odot$ \\
${\rm{M_{gas}/M_*}}$& 12.2\\
Total dynamical mass (${\rm{M_T}}$)&4.03 $\times~10^9$~M$_\odot$ \\
${\rm{M_T/L_B}}$& 149\\
${\rm{M_{dark}/M_{T}}}$(\%)& 95 \\
\hline
\end{tabular}
\end{table}

It is interesting to note that for a maximum halo circular speed 
of 50 kms$^{-1}$, comparable to the maximum velocity of NGC~3741, 
the scaling relations in \cite{bullock01} indicate that in an NFW type halo,
the rotation curve for NGC~3741 would start to decline at a radius of $\sim$ 7.6 kpc. 
This is just beyond the last measured point of our rotation curve. However the expected decline
is very gentle, and in the vicinity of 7.6 kpc is considerably smaller than the error in our rotation curve. 

\subsection{Star formation in NGC~3741}
\label{ssec:starform}

\begin{figure*}
\psfig{file=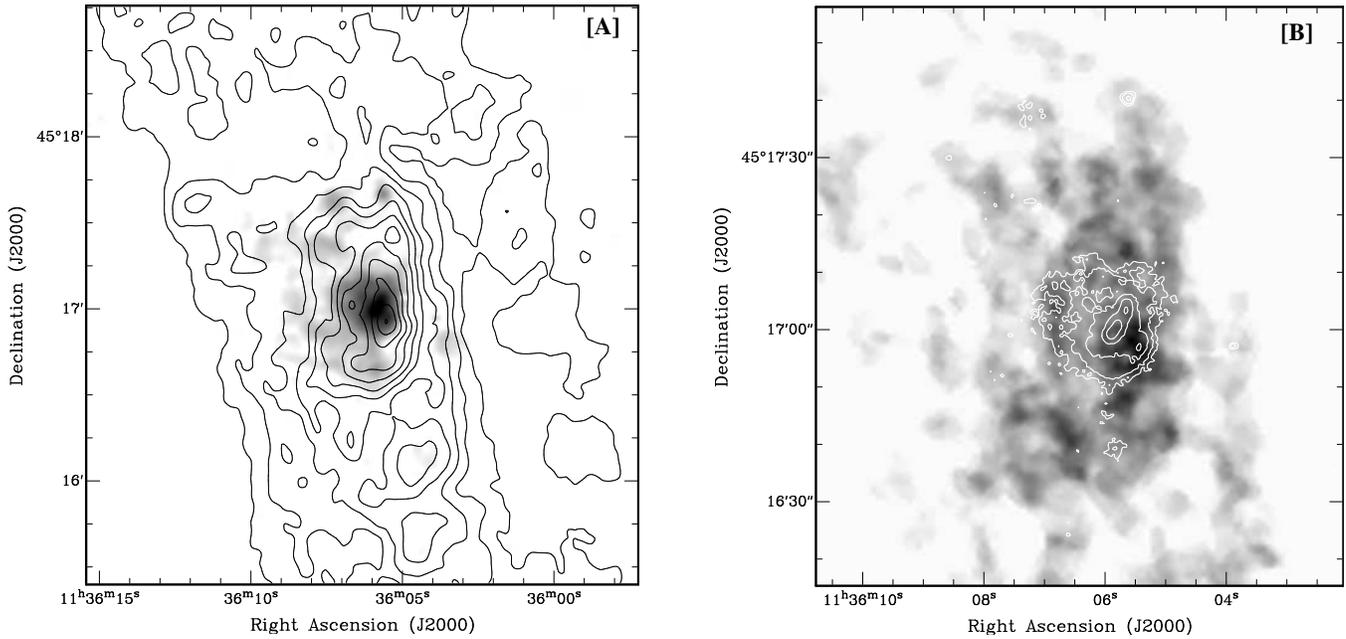,width=7.0truein,angle=-0}
\caption{[{\bf{[A]}}The integrated HI emission (contours) at a resolution of 11$^{''}\times9^{''}$, overlayed on the H$\alpha$ emission (greyscale). The contour levels start at 3.3 $\times 10^{20}$ cm$^{-2}$ and are uniformly spaced with a separation of
4.4 $\times 10^{20}$ cm$^{-2}$. The H$\alpha$ emission has been smoothed to a resolution of $3^{''}$
           {\bf{[B]}} Integrated HI image at a resolution of
4$^{''} \times 3^{''}$  (greyscale) overlayed on the H$\alpha$  emission (contours). The contour levels start at 4\% of the peak, and are spaced by a factor of 2. No smoothing
of H$\alpha$ emission is done.
}
\label{fig:halpha}
\end{figure*}

\begin{figure}
\psfig{file=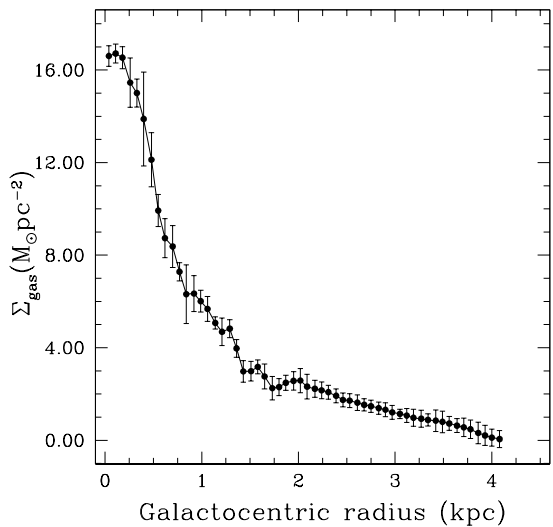,width=3.0truein,angle=-0}
\caption{The gas surface mass density profile for NGC~3741 derived from 11$^{''}\times9^{''}$
resolution map. The HI surface density profile has been scaled by 1.3 to account for the primordial
He.
}
\label{fig:smd_HI}
\end{figure}

\begin{figure}
\psfig{file=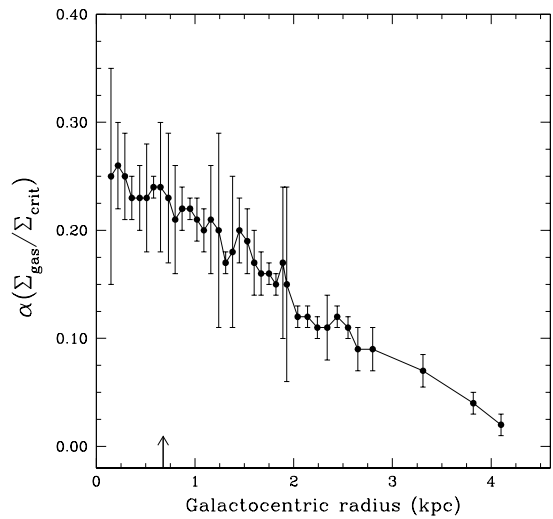,width=3.0truein}
\caption{The  ratio of HI surface density to the critical surface density from the Toomre's stability criterion, as a 
function of galactocentric radius.  The HI surface density profile is obtained
from 11$^{''}\times 9^{''}$ data cube (Fig.~\ref{fig:smd_HI}) and scaled by 1.3
to account for the primordial He fraction.
Errorbars take into account the difference between the forward and backward difference method
used to determine the derivative of the rotation curve for computing the critical surface density. 
The arrow indicates the Holmberg radius of the galaxy.
}
\label{fig:alpha}
\end{figure}

   As discussed in Section~\ref{ssec:rotcur}, NGC~3741 has one of the highest known ratios (for 
a gas-rich galaxy) of total mass to blue luminosity. On the other hand, as discussed in Paper~I, 
its ratio of total baryonic to total dynamical mass lies 
within the range that is typical for galaxies (Paper~I). 
In this section we use the HI and 
H$\alpha$ data to study the relationship between the star formation,  
gas distribution and gas kinematics in NGC~3741, in order to understand why the conversion
of gas into stars in this galaxy is extremely inefficient.

Figure~\ref{fig:halpha}[A] shows the HI column density distribution of NGC~3741 at a resolution 
of 11$^{''} \times 9^{''}$ (corresponding to a linear scale of $\sim$ 145 pc), 
overlayed on the H$\alpha$ emission (smoothed to a resolution of $3^{''}$) from the galaxy. 
A broad correspondence between the H$\alpha$ emission and high HI column density is seen in 
Figure~\ref{fig:halpha}[A], although the correspondence is far from being one to one.  
The contour which most closely matches the H$\alpha$ emission corresponds to an observed HI column density 
of $\sim 1.66\times 10^{21}$~cm$^{-2}$, 
or an inclination (and primordial He) corrected gas surface density of
$\sim 1\times10^{21}$~cm$^{-2}$. This is similar to the threshold value seen in other 
dwarf galaxies (e.g. \citealp{taylor94,skillman87}). Thus, most of the gas
disk of NGC~3741 is below the threshold column density required to induce star formation and 
this could be a reason why it was unable to convert most of its baryons 
into stars. However, what sets the threshold for star formation seen in NGC~3741?


The existence of a threshold surface 
gas density ($\Sigma_c$) for star formation in galactic disks is usually explained in terms of 
gravitational instabilities in thin rotating disk (i.e  Toomre's instability 
criterion; \cite{toomre64}, Kennicutt 1989), and is given as

\begin{equation}
\Sigma_{crit}=\alpha \frac{\kappa \sigma}{\pi G}
\label{eqn:sfr}
\end{equation}

where $\sigma$ is the velocity dispersion of the gas, $\kappa$ is the epicyclic frequency and $\alpha$
is a dimensionless constant included to account for a more realistic disk. 
In order to determine $\Sigma_{crit}$ for NGC~3741, we assumed $\alpha$ to be 1 initially
and determined its value from a comparison of the observed gas surface density and the threshold density.
Figure.~\ref{fig:alpha} shows the ratio of the  azimuthally  averaged gas surface density of NGC~3741, 
to the threshold  density predicted 
from Toomre's instability criterion.  The gas surface density profile for the 
galaxy (shown in Fig.~\ref{fig:smd_HI}) 
has been derived from 11$^{''}\times9^{''}$ resolution HI distribution and scaled by 1.3 to account for the 
primordial He fraction. No correction is made for the molecular gas. 
In deriving the threshold density, 
the velocity dispersion was assumed to be constant, and have a value of 8 \kms, a typical value observed 
for such faint dwarf galaxies (Begum et al. 2006).  As can be seen in Fig.~\ref{fig:alpha},  the 
ratio $\Sigma_{gas}/\Sigma_{crit} \sim 0.23$ for NGC~3741 within the star forming disk. 
For comparison, Kennicutt (1989) for a sample of bright Sc spiral galaxies found $\alpha \sim$0.67, assuming 
$\sigma$=6.0 kms$^{-1}$. Recomputing $\alpha$ for spirals (using Eqn.~\ref{eqn:sfr}),
using  $\sigma$ of 8.0 kms$^{-1}$ for NGC~3741 gives  $\alpha \sim$0.5. 
Thus, NGC~3741's star forming gas disk appears to be apparently 2 times more stable, 
compared to spiral galaxies, as per the Toomre's criterion.
This result suggest that, unlike spiral galaxies, Toomre's instability does not trace star formation in NGC3741.
This  is consistent with the behaviour seen in other dwarf galaxies, e.g. \cite{vanzee97, hunter98,begum06}. 
The exclusion of molecular gas in our analysis may have caused an 
underestimation of $\Sigma_{gas}/\Sigma_{crit}$. 
However, for Toomre's criterion to work for NGC~3741 requires as much molecular gas as
the atomic gas present within the star forming disk of this galaxy, which seems a priori unlikely. 
Sensitive observations of molecular gas in NGC~3741 are thus required to resolve this issue.
Further,  Toomre's stability analysis is based on the assumption that
the gas disk is thin, but this might not be the case for NGC~3741 as the outer 
gas disk is likely to be flaring in addition to being warped.
However, in the case of more  realistic thick disks,
the column density threshold is predicted to be higher 
by a factor of $\sim$ 1.5, as compared to thin disks ((Hunter et al. 1998, Romeo 1992).
Hence, if the gas disk of NGC~3741 is thick (which is more likely to be 
the case), it should be even more stable to star formation, compared to a thin disk case.

\begin{figure}
\psfig{file=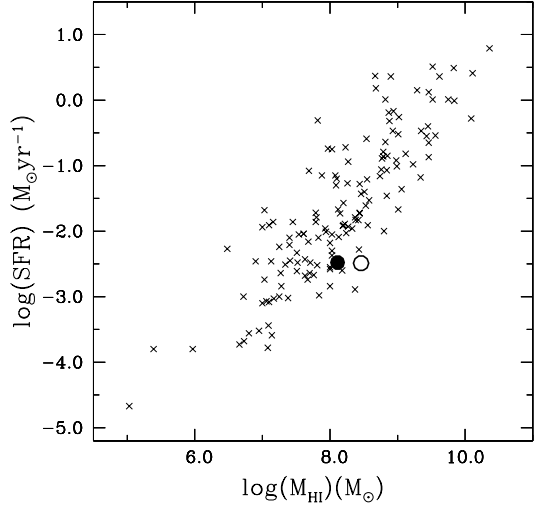,width=2.9truein,angle=0}
\caption{
Log of SFR plotted as a function of the HI mass for a sample of galaxies from Karachentsev \&
Kaisin (2007) (crosses). Gas rich galaxies with extended disks NGC~3741 (solid point) and DDO 154 (open circle)
are also shown. 
}
\label{fig:sfr_HI}
\end{figure}

The average star  formation rate of NGC~3741 
within the optical radius (R$_{25}$) is 4.86 $\times 10^{-3}~{\rm{M_\odot yr^{-1}kpc^{-2}}}$,
computed from the H$\alpha$ image, while the average HI mass surface density  within the same radius is
$\Sigma_{\rm{HI}}\sim15.0~{\rm{M_\odot pc^{-2}}}$.  These values are consistent with those expected
from the observed Kennicutt-Schmidt law for spiral galaxies (\citealp{kennicutt98}).
Further, Taylor \& Webster (2005) find that the star formation 
rate in dwarf galaxies correlates with the total HI mass, albeit with large 
(i.e. close to a factor of 10) scatter. Figure~\ref{fig:sfr_HI} shows the log of
SFR plotted as a function of the HI mass for a sample of galaxies from Karachentsev \& 
Kaisin (2007). NGC~3741 is shown as a solid point in the figure.
As seen in the figure, while the total HI mass of $1.3 \times 10^8$ M$_\odot$ (see Section~\ref{ssec:HIdistribution}) 
for NGC~3741 means that it has a lower SFR than the average for dwarf galaxies, 
the large scatter in the SFR-M$_{\rm HI}$ relation also means that it is not markedly deviant. 
In the SFR-M$_{\rm HI}$ plane, NGC~3741 lies close to DDO~154 (shown as a open circle), 
another gas-rich dwarf with a large HI envelope (\cite{carignan98}). 
Thus, we find that despite having a very extended gas disk, the global star formation properties 
of NGC~3741 are very similar to that of other dwarf galaxies.

Figure~\ref{fig:halpha}[B] shows HI image of NGC~3741  (greyscales) at 4$^{''} \times 3^{''}$ 
resolution (corresponding to a linear scale of $\sim 51$ pc), overlayed on the 
H$\alpha$ emission (contours). At this  
resolution the HI emission shows  substantial fine scale structure, with shell-like, 
filamentary and  clumpy features. One can also see that the 
H$\alpha$ emission is, by and large, offset from the high HI column density, with a general 
tendency for the H$\alpha$ to be surrounded by clumps of high  HI column density. This may be 
either because star formation has created a region of low local column density around it, or 
because the gas in these regions is either in an ionized or molecular phase.

\subsection{Chemical evolution of NGC~3741}
\label{sssec:chemevolution}

\begin{table}
\caption{Emission line fluxes, relative to H$\beta$, for HII region of NGC~3741 (from MK06) and derived abundance parameters.}
\label{tab:chemical}
\vskip 0.1in
\begin{tabular}{ll}
\hline
Parameters& Value \\
\hline
\hline
[OII] $\lambda$3727 & 1.67 $\pm$0.20\\
H$\beta$ (4861 \AA) & 1.00$\pm$ 0.07\\
$\rm{[OIII]}$ $\lambda$5007 & 1.68$\pm$0.11\\
H$\alpha$ (6563 \AA) & 2.85$\pm$0.18\\
$\rm{[NII]}$ $\lambda$6584 & 0.05$\pm$ 0.20\\ 
F(H$\beta$) (ergs s$^{-1}$ cm$^{-2}$) & 123.7 $\times 10^{-15}$\\
C(H$\beta$) & 0.00$\pm$0.18\\
log(R$_{23}$) & 0.59 $\pm$0.07\\
log([OIII])/[OII]) &0.13$\pm$0.11\\
log(O/H) &7.66$\pm$0.10\\
\hline
\end{tabular}
\end{table}

In this section we estimate the HII region (oxygen) abundance of NGC~3741 and study its chemical enrichment in light
of the closed-box chemical evolution in galaxies.
NGC~3741 has been observed as a part of a spectrophotometric survey of nearby star forming galaxies (Moustakas \& Kennicutt, 
2006, henceforth MK06). For the details of the spectrophotometric observations and analysis the interested readers are referred to MK06 paper. The integrated  emission-line fluxes relative to H$\beta$, corrected for foreground Galactic extinction and for
underlying stellar absorption, as obtained by MK06 are listed in Table~\ref{tab:chemical}.  To correct the observed line ratios for reddening,  the intrinsic case B Balmer line ratios were taken from Osterbrock (1989) assuming n$_e=100 {\rm{cm}}^{-3}$ and T$_e=10^4$ K.
The value of the logarithmic extinction (or reddening coefficient) at H$\beta$, C(H$\beta$) was hence  derived from the H$\alpha$/H$\beta$ ratio. The estimated reddening for NGC~3741 was found to be negligible (C(H$\beta$)=2.6 $\times 10^{-3}$). 

No [OIII]$\lambda$4363 was detected in NGC~3741, hence 
in the absence of the measurement of I([OIII]$\lambda4363$), the oxygen abundance of the galaxy was estimated using the
bright emission lines [OII] and [OIII], using a grid of photo-ionization models developed by McGaugh (1991). In such
models R$_{23}$=[I([OII])+I([OIII])]/I(H$\beta$) and [OIII]/[OII] ratios are used to derived an empirical estimate of the oxygen abundance.  As R$_{23}$ is double valued, the degeneracy of the R$_{23}$ relation was broken by inspecting the value of [NII]/[OII] ratio (van Zee et al. 1998). For NGC~3741 log([NII]/[OII])$\sim - 1.5$, hence the low-abundance branch of the model grid of the R$_{23}$ relation from McGaugh (1991) was considered. The derived ratios for NGC~3741 are listed in Table~\ref{tab:chemical}. The oxygen abundance derived using the bright line calibration was hence found to be  12+ log(O/H)= 7.66$\pm$0.10. The uncertainty in this estimate also includes the uncertainty in the model calibrations of the semiemperical
relation between line strength and elemental abundance for log(O/H) (McGaugh 1991).

\begin{figure}
\psfig{file=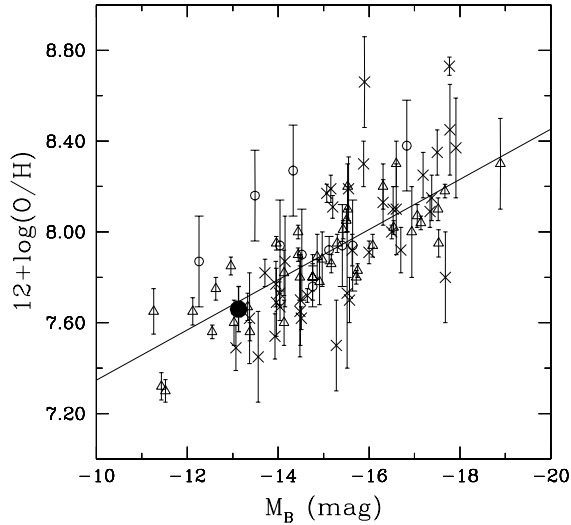,width=3.0truein}
\caption{Metallicity-luminosity relation for a sample of galaxies. Open circles are from Lee et al. (2007), crosses from
van Zee \& Haynes (2006) and open triangles from van Zee et al. (2006). NGC~3741 is shown as a solid point. Solid line
shows the best fit metallicity-luminosity relation to all the data.
}
\label{fig:metal-lumin}
\end{figure}

Figure~\ref{fig:metal-lumin}  shows the absolute B magnitude for NGC~3741, plotted against the derived oxygen abundance.  The same quantity
is also plotted for a sample of galaxies compiled from literature;   
the sample from which these galaxies have been drawn are listed in the figure caption. The solid line with the slope and intercept of 
\linebreak
  ( $-$0.11$\pm$0.01, 6.24$\pm$0.16) respectively, shows the
best fit to the data. As can be seen NGC~3741 follows the metallicity-luminosity relation as defined by other dwarf galaxies.   


We will now compare the derived oxygen abundance and gas fraction of NGC~3741 with expectations from the closed-box chemical 
evolution in galaxies. In a closed-box model the baryonic gas mass fraction $\mu$ 
(the ratio of the gas mass to the total mass in gas and stars) and the gas-phase 
abundance (Z) are related as 

\begin{equation}
{\rm{Z=p~ln}}(1/\mu)
\label{eqn:closedbox}
\end{equation}

{\hskip-0.70cm}where p is the elemental yield by mass. The effective yield, p$\rm{_{eff}}$, is defined as 
the yield that would be deduced if the galaxy was assumed to be a simple closed-box.
In order to compute $\mu$ for estimating p$\rm{_{eff}}$, we need to know the amount of gas 
that could participate in star formation. For galaxies with extended HI disks 
whether to compute $\mu$ using the entire HI mass or just the HI mass within the optical disk is 
an important issue (Garnett 2002). We computed p$\rm{_{eff}}$ for the
 two limiting cases  i.e. (i) using the entire HI mass and (ii) using only the HI mass within the optical disk.
The effective yield of oxygen for NGC~3741, using the entire HI mass,
is found to be p$\rm{_{eff}}=0.0069 \pm 0.0007$ (derived using the parameters in Table.~\ref{tab:par}). 
In the model used by vanZee \& Haynes (2006) (henceforth VH06), this yield is 
consistent (within the errorbars) with the  theoretically 
expected closed-box yield  of p=0.0074 (Meynet \& Maeder, 2002). On the other hand, using only the
HI mass within the optical disk, the derived p$\rm{_{eff}}\sim 7.5 \times 10^{-4}$ is much lower than that expected from
closed-box chemical evolution.


\begin{figure}
\psfig{file=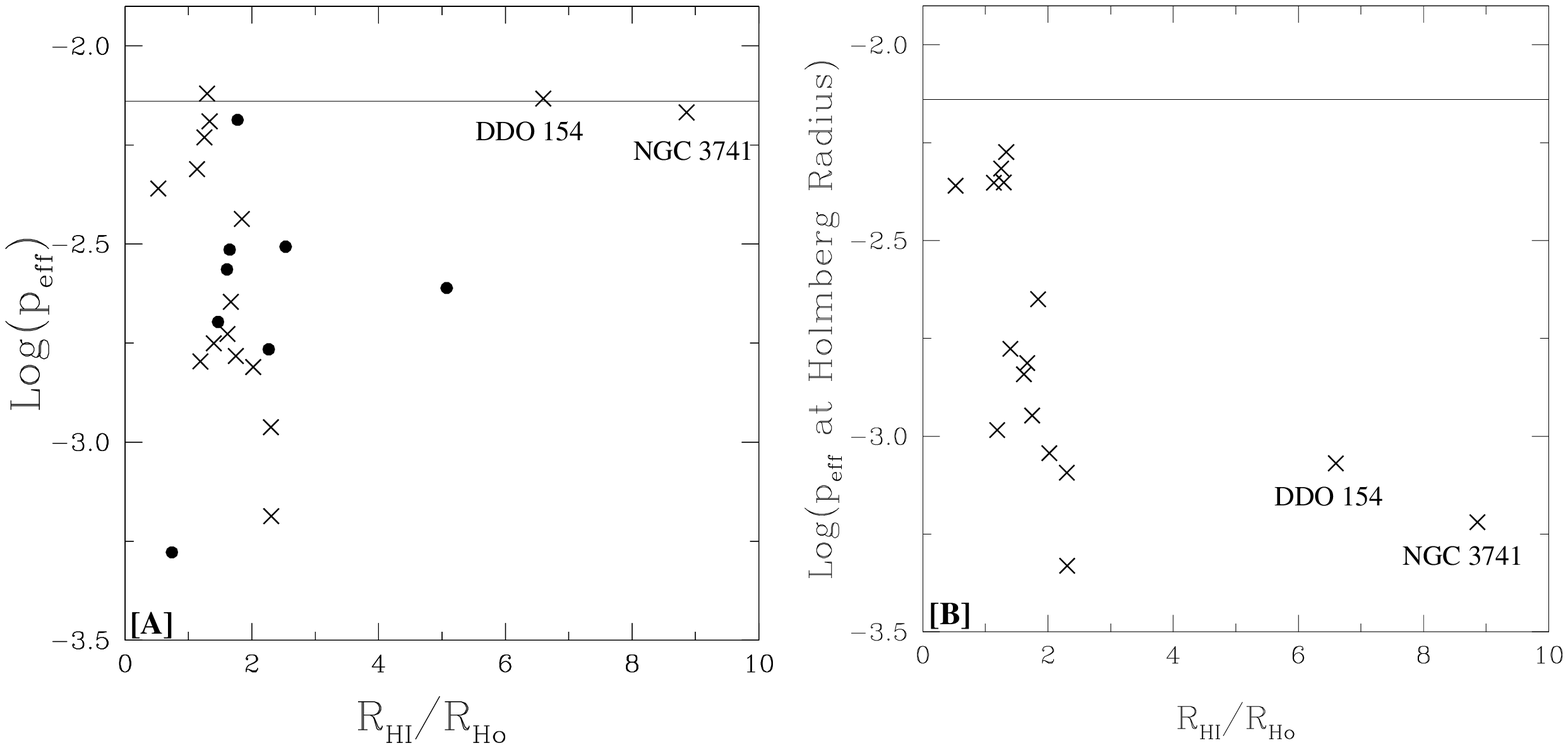,width=3.5truein}
\caption{{\bf{[A]}}The effective yield for a sample of galaxies (from VH06 sample for which interferometric HI images are available) as 
a function of the HI extent of the galaxy. The effective yield is computed from the gas mass within the total HI extent of the galaxy. 
Solid points indicate the galaxies for which only the HI extent is available 
in literature but not the detailed HI distribution. The sources of HI images 
van Zee et al. 1997, Begum et al. 2006, Begum \& Chengalur 2003, Swaters et al. 2002.
{\bf{[B]}} Effective yield is computed 
using the gas mass within the Holmberg radius. The solid line shows the theoretically
expected closed-box yield of p=0.0074 (Meynet \& Maeder, 2002).
}
\label{fig:overlay_yield}
\end{figure}

\begin{figure}
\psfig{file=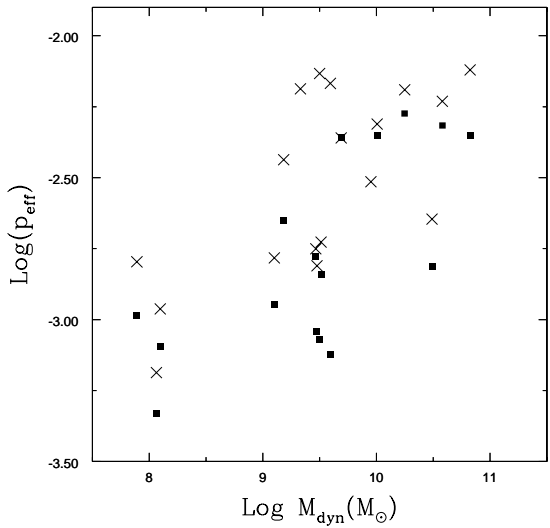,width=3.0truein}
\caption{The effective yield for a sample of galaxies (from VH06 sample for which interferometric HI images are available),
computed from the gas mass within the total HI extent of the galaxy (crosses) and only from the HI extent within
the Holmberg diameter (solid points), as a function of the dynamical mass of the galaxy. The sources of
the HI data is same as in Fig.~\ref{fig:overlay_yield}.
}
\label{fig:yield_mass}
\end{figure}

If interpreted literally, the above exercise would suggest that  NGC~3741 is evolving as a closed-box model, 
provided that the gas-phase metallicity in the inner disk is the same as the outer disk, i.e. that there 
is an efficient mixing of the metals throughout the HI disk. However this seems unlikely, given the 
large size of the HI disk. 
Tassis et al. (2006) showed that the mixing length of metals in a galaxy increases with a 
decrease in the galaxy mass. However, whether there is mixing of metals  even up to 14 times its optical 
extent is unclear.
On the other hand, what evidence do we have for closed-box chemical evolution in dwarf galaxies ?
And can one from the observational data try to make inferences about how much of the HI disk participates
in the chemical evolution (i.e. is well mixed)?
The gravitational binding energy of faint dwarf irregular galaxies is not much larger than the energy output
from a few supernovae, hence a priori, one might expect that low mass galaxies would
depart from the closed-box chemical evolution, since enriched material could
escape via stellar winds and supernova ejecta (Brooks et al. 2007). 
We searched the literature for the HI interferometric images for the galaxies 
in VH06 sample and show in Fig.~\ref{fig:overlay_yield}[A]  the effective yield 
plotted as a function of the HI extent of a sample of galaxies in VH06 sample with 
available HI images (the sources of the HI images is given in the figure caption). 
The effective yield is computed by considering the entire gas mass. NGC~3741 and 
DDO154 are shown in the plot with extreme values of the HI extent. As can be seen, 
there are some galaxies in the sample which are consistent with the closed-box model.
However there is no clear trend  seen between the effective yield and the extent  of the HI disk.  
Figure~\ref{fig:overlay_yield}[B] shows the effective yield for the same sample of galaxies as shown in 
Fig.~\ref{fig:overlay_yield}[A], however this time the effective yield is computed by considering the gas fraction within the
Holmberg radius. As can be seen, if the effective yield is computed
within the Holmberg radius, (as would be appropriate in the case
of inefficient mixing) none of these galaxies follow 
a closed-box model.


The theoretically expected closed-box chemical yield depends on
assumptions about the IMF, stellar rotation as well as in details of stellar evolution 
models (Meynet \& Maeder, 2002). It may be more instructive hence 
to look for trends in the effective yield as a function of other galaxy properties, 
instead of comparing the observed yield against a model dependent expected yield. 
One possible parameter to use for such a correlation is the total dynamical mass.
VH06 did not find any significant trend between the dynamical mass
and effective yield for the galaxies in their sample, however they 
computed the dynamical mass using the HI global velocity widths. We recomputed the 
dynamical masses for the galaxies in VH06 sample
for which rotation curves are available in literature. We show in 
Fig.~\ref{fig:yield_mass} a plot
of the effective yield against the total dynamical mass. The effective
yield computed using the entire HI mass is shown as cross, while
the effective yield computed using the HI mass within the Holmberg
radius is shown as solid points. As can be seen, in both cases the
effective yield increases with increasing dynamical mass.  The correlation 
is tighter if one uses the only HI mass within the Holmberg radius (correlation
coefficient is 0.59) compared to using the entire HI mass (correlation coefficient is 0.47), 
suggesting that, if this model were to be correct, only the gas within 
the optical disk participates in the chemical evolution of the galaxy.
However we note that the total number of galaxies in our analysis is too
small to make any statistical conclusion.
Spectroscopic observations of a large sample of dwarf galaxies along with
a knowledge of the gas distribution is hence required
for a better understanding of chemical enrichment and mixing of
enriched material in gas-rich, low mass galaxies.

   In summary, we examine the dark and luminous matter in NGC~3741. Although this galaxy has one of the 
highest known ratios of dark to luminous (i.e. stellar) matter, its baryons to dark matter
ratio is typical of that of galaxies. In its global star formation
properties however, the galaxy appears to follow the trends defined
by other dwarf galaxies. Given the large observed scatter about these mean trends, it is perhaps not very surprising that NGC~3741 is not
very markedly discrepant. Finally we examine the issue of chemical
evolution via a closed-box model. The
effective oxygen yield of NGC~3741 is that expected from theoretical
models, provided one assumes that there is efficient mixing 
of the gas $\sim 8$ times further than the end of the optical disk.
This seems apriori unlikely. On the other hand,
using a sample of galaxies with both interferometric HI maps and chemical
abundance measurements, we find that the effective yield is anti-correlated with the total
dynamical mass, as expected in leaky box models.

\section*{Acknowledgments}
We are very thankful to Prof Liese van Zee for discussions regarding the effective yield of dwarf galaxies. 
We thank Dr. G. Gentile for providing the rotation curve of NGC~3741 derived from
the WSRT data. Thanks to Dr Christy Tremonti and Dr Shoko Sakai for useful comments on this paper.
Partial support for this work was provided by ILTP grant B-3.13.
The Westerbork Synthesis Radio Telescope is operated by the ASTRON (Netherlands Foundation for 
Research in Astronomy) with support from the Netherlands Foundation for Scientific Research NWO.
The GMRT is operated by the National Center for Radio Astrophysics of the Tata Institute
of Fundamental Research.

\end{document}